# NMR diagnosis of pseudo-scalar superconductivity in 3D Dirac materials

Azin Mohajerani[1], Zahra Faraei[2] and S. A. Jafari[2,3]

[1] *Department of Basic Sciences, Tarbiat Modares University (TMU), Tehran14115-175, Iran*
[2] *Department of physics, Sharif University of Technology, Tehran 11155-9161, Iran*
[3] *Centre of excellence for complex systems and condensed matter (CSCM), Sharif University of Technology, Tehran 1458889694, Iran.*

Recently observed $4\pi$ periodic Andreev bound states in three dimensional Dirac materials are attributed to convnetional superconducting pairing. Our alternative explanation in terms of a novel form of parity breaking pseudo-scalar superconducting order can be sharply diagnosed by nuclear magnetic resonance (NMR) relaxation rate. The left-right symmetry breaking of the pseudo-scalar superconductivity can be directly probed as an anti-peak structure below $T_C$ in sharp contrast to the conventional Hebel-Slichter peak.

## 1. Introduction

Recently discovered three dimensional Dirac materials (3DDMs) - $Na_3Bi$ [1,2], $Bi_{1-x}Sb_x$ [3,4] and $Cd_3As_2$ [5,6] - have interesting properties due to their bulk linear energy dispersion. They have fourfold degeneracy at Dirac point and can be described by 3D massive Dirac Hamiltonian [7,8].

Superconducting state can be achieved by applying pressure [9], using point contacts [10, 11] or proximity effect [12, 13, 14] in 3DDMs. In the proximity of a 3DDM to a conventional s-wave superconductor, an unconventional multicomponent order parameter will be induced. Chuan Li, et al. [12] have used nanoscale phase-sensitive junction technology to induce superconductivity in $Bi_{1-x}Sb_x$ and demonstrated a significant contribution of $4\pi$-periodic Andreev bound states to the supercurrent in Nb-$Bi_{0.97}Sb_{0.03}$-Nb Josephson junctions by subjecting it to radio frequency irradiation. The $4\pi$-periodic current phase relation is a clear signature of the Majorana zero modes [15, 16]. The authors propose that the Klein tunnelling mechanism of the incident electrons on the superconductor gives rise to perfect transmission which in turn results in 100% Andreev reflection guaranteeing the appearance of Majorana zero modes. However, the topological protection against backscattering in Dirac cone in the 3DDM interlayer takes place only for discrete incident angles. Such angles form a set of measure zero and therefore are not expected to lead to observable signal. Similarly, $\pi$ and $4\pi$-periodic supercurrents is observed in Al-$Cd_3As_2$-Al Josephson junctions[17]. They propose the $\pi$-periodic supercurrent state arising from the coupling of induced Fermi arc surface superconductivity and bulk normal pairing.

Alternative explanation can be put forward in terms of novel forms of superconducting order which is permitted by the symmetries of Dirac equation. Recent theoretical studies of induced superconductivity in a 3DDM using proximity effect [13] suggest that a simple conventional BCS superconductor when proximitiezed with a Dirac material, among others allow for parity breaking superconducting order. To formalize this, take a 3D Dirac Hamiltonian of a system with strong spin-orbit coupling which can be expressed in terms of $4\times 4$ Dirac $\gamma$ matrices such that the electron wave funcitons are spinors with four components. Therefore to Cooper pair them, there are 16 possibilities which can be incoded into a $4\times 4$ matrix. This matrix on the other hand can be decomposed in a basis of matrices with definite transformation properties under Lorentz symmetry. Marices that involve $\gamma^5$ break an additional left-right symmetry in the chirality space, meaning that under parity of mirror reversal they change sign. This form of *pseudo-scalar* superconductivity is spin-singlet [13].

It was noticed [14] that the Dirac-Boguliubov-de Gennes equation with the superconducting pairing matrix proportional to $\gamma^5$ is topologically non-trivial and belongs to DIII class and hence is classified by integer topological index. This gaurantees not only a topological protection for the Andreev zero mode, but also gives rise to 100% Andreev reflection for *any* electron imping on the superconductor. This is in sharp contrast to the scalar (BCS) induced superconducting order where only at some accidental angles of incidence the 100% Andreev reflection can be achieved [12].

Nuclear magneitc resonance (NMR) relaxation rate [18,19] as a bulk quantity sensitive to symmetry of the gap function is a practical probe to identify the various superconducting states. It is a powerful tool to differentiate between conventional and exotic superconductivity. A classical exmaple is the distinction between singlet and triplet pairing by NMR [20] where the difference in the symmetry of singlet and triplet superconducting order parameter can be detected via Knight shift and spin relaxation rate $T^{-1}$ measurements. Now the question is, given that both scalar and pseudo-scalar superconducting orders are possible in a 3DDM, can NMR dignose the breaking of left-right symmetry of the pseudo-scalar order? The answer we find in this work is affirmative. NMR can identify a key difference between the scalar and pseudo-scalar superconductors. In this work we reveal that the spin singlet pseudo scalar superconducting state with even momentum and odd mirror parity induced in a 3DDM can be identified through the tempereture dependence of the nuclear magnetic relaxation rate. We find that unlike the conventional superconduting case, the pseudo-scalar superconductivity gives rise to an anti-peak structure below the transition temperature.

## 2. Model and Methods

The minimal model for a superconducting 3DDM is Dirac Bogoliubov-de-Gennes Hamiltonian [21]:

$$\breve{H}(k) = \begin{bmatrix} \hat{H}_{0D}(k) & \hat{\Delta}(k) \\ \hat{\Delta}^\dagger(k) & -\hat{H}_{0D}(k) \end{bmatrix}. \quad (1)$$

The notation (˘) stands for matrices in Nambu space, and the (ˆ) indicates the matrix structure in orbital-spin space. The normal-state Hamiltonian matrix describes massive Dirac electrons,

$$\hat{H}_{0D}(k) = m\gamma^0 + k_i \gamma^0 \gamma^i - \mu, \quad (2)$$

with chemical potential $\mu$ and mass $m$. Here $\gamma^i$ ($i=1,2,3$) are $4 \times 4$ Dirac $\gamma$ matrices[22]:

$$\gamma^0 = \tau_3 \otimes \mathbb{1}, \qquad \gamma^i = \tau_1 \otimes i\sigma_i,$$

where $\tau_i(\sigma_i)$ denote the $2 \times 2$ Pauli matrices in orbital (spin) space.

In proximity of a conventional s-wave superconductor with a 3DDM, Green's function method and tunnelling formulation give rise to an uconventional multicomponenet order parameter. The Fierz decomposition [23] of the superconducting matrix will be

$$\hat{\Delta} = \Delta_s \mathbb{1} + \Delta_\mu \gamma^\mu + \Delta_{\mu\nu}\sigma^{\mu\nu} + \Delta_{5\mu}\gamma^{5\mu} + \Delta_5 \gamma^5, \quad (3)$$

where $\sigma^{\mu\nu} = i\gamma^\mu \gamma^\nu$ ($\mu,\nu = 0,1,2,3$). Apparently, the induced superconductivity in a 3DDM has sixteen possible channels categorized as scalar $\Delta_s$, (four) vector $\Delta_\mu$, tensor $\Delta_{\mu\nu}$, pseudo vector $\Delta_{5\mu}$ and pseudo scalar $\Delta_5$ states. All these various channels have different spin, momentum and orbital symmetry which are not independent. They must fullfill the fermionic anticommutation property of the pairing potential. The scalar, pseudo scalar order paramerters have both even momentum dependence and are spin-singlet, while the matrix structure in the space of Lorentz indices (spin and orbital space) are different. As indicated in Eq. (3), the scalar order comes hand in hand with unit $4 \times 4$ matrix, while the pseudo-scalar order is associated with the $\gamma^5$ matrix. In the following we show how $\gamma^5$ matrix unlike any other matrix in Eq. (3) generates an extra minus sign in NMR relaxation rate, which will eventually lead to an anti-peak structure.

The nuclear spin-lattice relaxation rate for a two band superconductor[24, 25] is given by,

$$\frac{1}{T_1 T} = \pi \sum_{\alpha\acute\alpha} \int_{-\infty}^{\infty} d\omega \left[-\frac{df(\omega)}{d\omega}\right] \\ \times Re\left\{S_{\uparrow\uparrow}^{G\alpha\acute\alpha}(\omega) S_{\downarrow\downarrow}^{G\acute\alpha\alpha}(\omega) - S_{\uparrow\downarrow}^{F\alpha\acute\alpha}(\omega)\left[S_{\downarrow\uparrow}^{F\acute\alpha\alpha}(\omega)\right]^*\right\}. \quad (4)$$

The indices α and α′ represent band labels. The $f(\omega)$ is the Fermi-Dirac distribution function. $S^G$ and $S^F$ are spectral representation of the normal and anomalous Green's functions and are given by,

$$S^{G(F)}(\omega) = \frac{-1}{2\pi i}\sum_k [G(F)_k(i\omega_n \to \omega + i\delta) - G(F)_k(i\omega_n \to \omega - i\delta)], \quad (5)$$

with fermionic matsubara frequency $\omega_n = \pi T(2n+1)$ $n \in \mathbb{Z}$ and

$$\hat{G}(k,\omega_n) = -\frac{i\omega_n + \xi(k)}{\omega_n^2 + E_k^2}\sigma_0,$$

$$\hat{F}(k,\omega_n) = \frac{\hat{\Delta}(k)}{\omega_n^2 + E_k^2}. \quad (6)$$

The Hebel-Slichter peak just below $T_C$ in a superconductor stem from the seconed term in Eq. (4). Obviously, the anomalous spectral function only contributes to the peak if the order parameter has even momentum dependence. Moreover, the sign in front of the second term accounts for the enhacement behavior in NMR rate below $T_C$. To see this, for simplicity assume that there are no band indices (i.e. we are dealing with one-band situation). In this case a minus sign arising from the spin-exchange in the spin-singlet pairing generates an overal positive sign before the second term multiplied by the modulus of a $S^F$ term which is a positive definte quantity. This is how the celebrated Hebel-Slichter peak is generated for conventional spin singlet s-wave superconductors. Now for scalar pairing, the band indices do not harm the above argument, and the band indices are diagonally identified, and we end up with sum of four positive defnite contributions to the second term of Eq. (4). This argument applies to all other matrices appearing in the Fierz decomposition Eq. (3), except for $\gamma^5$ alone. In the following let us see how this happens.

Among the 16 possible superconducting pairing channels induced in the 3DDM, $\Delta_s$, $\Delta_0$, $\Delta_5$ and $\Delta_{50}$ have even momentum dependence and hence survive the momentum integration. Therefore these terms have non-zero contributions to the coherence effect in NMR rate. All odd momentum dependence have null contribution under momentum integration. Hence we are left with,

$$\begin{aligned}
\hat{\Delta}_s &= \Delta_s \mathbb{1}, \\
\hat{\Delta}_0 &= \Delta_0 \gamma^0 & \gamma^0 &= \tau_z \otimes \mathbb{1}, \\
\hat{\Delta}_5 &= \Delta_5 \gamma^5 & \gamma^5 &= -\tau_y \otimes \mathbb{1}, \\
\hat{\Delta}_{50} &= \Delta_{50} \gamma^{50} & \gamma^{50} &= i\tau_x \otimes \mathbb{1}.
\end{aligned} \quad (7)$$

Since these four pairing potentials are odd under the spin exchange [13], the coherence term in Eq. (4) does change the sign for all of them. The structure of the second term of Eq. (4) in the space of Lorentz indices is of the form $\text{Tr}(\gamma\gamma^*)$ as follows:

$$T_1^{-1} \sim - S_{\uparrow\downarrow}^{F\alpha\acute{\alpha}}(\omega) \left[S_{\uparrow\downarrow}^{F\acute{\alpha}\alpha}(\omega)\right]^\dagger \sim -\Delta_{\downarrow\uparrow}^{\alpha\acute{\alpha}}(k,\omega) \left[\Delta_{\uparrow\downarrow}^{\alpha\acute{\alpha}}(k,\omega)\right]^\dagger$$
$$\sim \Delta_{\uparrow\downarrow}^{\alpha\acute{\alpha}}(k,\omega) \Delta_{\downarrow\uparrow}^{*\acute{\alpha}\alpha}(k,\omega).$$

From the explicit representation of gamma matrices in Eq. (7) one can easily check that among the above matrices, after the multiplication with its complex conjugate matrix, only $\gamma^5$ gives rise to an *extra minus sign*.

## 3. Disscussion

The sign reversal associated with pseudo-scalar pairing relies on two facts: (i) odd mirror parity of the pseudo-scalar pairing. (ii) The complex conjugation inherited from the condensate structure of the anomalous Green's functions. Indeed the complex conjugation rests on the very basic facts that the Cooper pairs carying charge 2e are constructed from a given state and its charge conjugate which then brings in the complex conjugation. This is therefore a unique ability of NMR that can probe a combined effect of complex conjugation and mirror-parity. Hence, the NMR relaxation rate is able to realize the breaking of mirror symmetry (as it happens in pseudo scalar superconductor).

Note that this is a unique capability of NMR as compared to closely related probes such as ultrasonic attenuation and electromagnetic absorption. Since the ultrasonic and electromagnetic absorption rates do not directly couple to condensate, they are not sensitive to left-right breaking character of the pseudo-scalar order parameter.

## 4. conclusion

In this work we considered the sixteen possible superconducting order parameters in a three dimensional Dirac material. The generic superconducting order encoded in a $16 \times 16$ matrix can be decomposed into various channels with definite transformation properties under rotations. Among them a topologically non-trivial superconducting order is the pseudo-scalar one. This corresponds to a pairing order parameter that changes sign in the mirror. The non-trivial topology of this pairing protects the ensuing Majorana zero modes. These Majorana zero energy states are responsible for the 100% Andreev reflection at zero energy which is equivalent to $4\pi$ periodic Andreev bound states observed in superconducting 3DDM [12].

Our explanation of the $4\pi$ periodic Josephson effect in terms of pseudo-scalar superconductivity [13] can be validated by an anti-peak structure in the NMR relaxation rate.

## Acknowledgements

A. Mohajerani is grateful to Dr. Y. Fathollahi for his help and wisdom that propelled the creation of this research.